 \documentclass[12pt]{article}
 \usepackage {graphicx}
 \usepackage{epsfig}
 \usepackage[latin1]{inputenc}
 \usepackage{amsmath}
 \usepackage{amsfonts}
 \usepackage{amssymb}
 \usepackage{amsthm}
 \newcommand{\be}{\begin{equation}}
 \newcommand{\ee}{\end{equation}}
 \newcommand{\ba}{\begin{eqnarray}}
 \newcommand{\ea}{\end{eqnarray}}
 \newcommand{\inc}{{\it{i}}}

 \newcommand{\efbold}{\mbox{{\boldmath $\vec f$}}}
 
 \newcommand{\erbold}{\mbox{{\boldmath $\vec r$}}}
 
 \newcommand{\omegabold}{\mbox{{\boldmath $\vec \omega$}}}

 \newcommand{\vbold}{\mbox{{\boldmath $\vec v$}}}
 \newcommand{\Vbold}{\mbox{{\boldmath $\vec V$}}}
 \newcommand{\Hbold}{\mbox{{\boldmath $\vec H$}}}

 \newcommand{\fbold}{\mbox{\boldmath $\vec {\boldmath{\,f}}$}}
 
 \newcommand{\efcal}{\mbox{{\boldmath $\vec{\cal{F}}$}}}
 
\begin{document}
 \title{
    ${{~}^{^{^{Published~in~the~Journal~of~Geophysical~Research\,-\,Planets\,,
    ~Vol.\,112\,,~p.\,E12003~\,(2007) }}}}$\\
 ${{~~~~~~~~~~~~~~~~~~~~~~~~~~~~~~~~~~~~~~~~~~~~~~~~~~~~~~~~}^{^{^{{doi}:~10.1029/2007JE002908 }}}}$\\
 {\Large{\textbf{The Physics of Bodily Tides in Terrestrial Planets,\\
                 and the Appropriate Scales of Dynamical Evolution\\}
            }}}
 \author{
 {\Large{Michael Efroimsky}}\\
 {\small{US Naval Observatory, Washington DC 20392 USA}}\\
 {\small{e-mail: ~me @ usno.navy.mil~}}\\
 ~\\
 {and}\\
 ~\\
 {\Large{Val{\'e}ry Lainey}}\\
 {\small{IMCCE-Observatoire de Paris, UMR 8028 du CNRS, ~77 Avenue Denfert-Rochereau,
   ~Paris ~75014 ~France,}}\\
{\small{Observatoire Royal de Belgique, ~3 Avenue Circulaire, ~Bruxelles ~1180
   ~Belgique}}\\
 {\small{e-mail: ~lainey @ imcce.fr~~~~~~~~}}
 }
 \date{}
 \maketitle
 \begin{abstract}
 \noindent
 Any model of tides is based on a specific hypothesis of how lagging depends on
 the tidal-flexure frequency $\chi$. For example, Gerstenkorn (1955), MacDonald
 (1964), and Kaula (1964) assumed constancy of the geometric lag angle $\,\delta\,
 $, while  Singer (1968) and Mignard (1979, 1980) asserted constancy of the time
 lag $\,\Delta t\,$. Thus, each of these two models was based on a certain law of
 scaling of the geometric lag: the Gerstenkorn-MacDonald-Kaula theory implied that
 $\,\delta\,\sim\,\chi^0\,$, while the Singer-Mignard theory postulated $\,\delta\,
 \sim\,\chi^1$.

 The actual dependence of the geometric lag on the frequency is more
 complicated and is determined by the rheology of the planet. Besides, each
 particular functional form of this dependence will unambiguously fix the
 appropriate form of the frequency dependence of the tidal quality factor,
 $\,Q(\chi)$. Since at present we know the shape of the function $\,Q(\chi)\,$,
 we can reverse our line of reasoning and single out the appropriate actual
 frequency-dependence of the lag, $\,\delta(\chi)\,$: as within the frequency
 range of our concern, $\,Q\,\sim\,\chi^{\alpha}\,,\;\alpha\,=\,0.2\,-\,0.4\,$,
 then $\,\delta\,\sim\,\chi^{-\,\alpha}\,$. This dependence turns out to be
 different from those employed hitherto, and it entails considerable alterations
 in the time scales of the tide-generated dynamical evolution. Phobos' fall on
 Mars is an example we consider.
 \end{abstract}

 \section{Introduction}

\begin{figure}
 \includegraphics[height=.42\textheight,
 ]{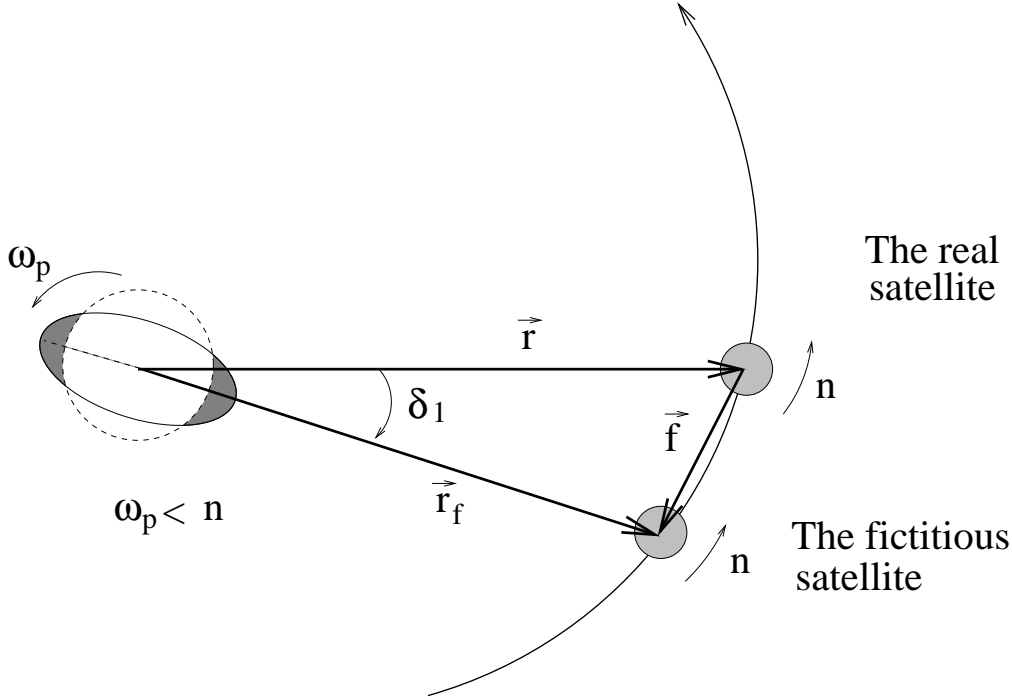}
 \caption{\small
 A planet and a tide-raising moon. This picture illustrates the case of a
 satellite located below the synchronous orbit, so that its mean motion $\,n\,$
 exceeds the planet's spin rate $\,\omega_p\,$, and the tidal bulge
 is lagging. The angular lag defined as $\,\delta\,\equiv\,
 {\textstyle{|\efbold|}}/{\textstyle{r}}\,=\,\frac{\textstyle\Delta t}{
 \textstyle{r}}\,|\,\omegabold_p
 \times\erbold\;-\;\vbold\,|\,$ will, generally, differ from the absolute value of
 the angle $\,\delta_1
 \,$ subtended at the planet's centre between the directions to the satellite and
 the bulge. Since in our study we consider an example with a small eccentricity and
 inclination, we make no distinction between
 $\,\delta\,$ and $\,|\delta_1|\,$.
 }
 \label{fig-JGR.eps}
 \end{figure}

 If a satellite is located at a planetocentric position $\erbold$, it generates a
 tidal bulge that either advances or retards the satellite motion, depending on the
 interrelation between the planetary spin rate $\omega_p$ and the tangential part of
 satellite's velocity $\vbold$ divided by $r\equiv|\erbold|$. It is convenient to
 imagine (as on Fig.~\ref{fig-JGR.eps}) that the bulge emerges beneath a fictitious
 satellite located at
 \ba
 \erbold_f\,=\,\erbold\,+\,\fbold\;\;\;,
 \label{1}
 \label{401}
 \ea
 ~\\

 \noindent
 where the position lag $\,\fbold\,$ is given by
 \ba
 {\vec{\mbox{\it\textbf{f}}}}\;=\;\Delta t\;\left(\;\omegabold_p\times
 \erbold\;-\;\vbold\;\right)\;\;\;.
 \label{2}
 \label{402}
 \ea
 $\Delta t\,$ is the time lag between the real and fictitious
 tide-generating satellites, and the inclination and eccentricity of the satellite
 are assumed sufficiently small.

 The fictitious satellite is merely a way of illustrating the time lag
 between the tide-raising potential and the distortion of the body. This
 concept implies no new physics, and is but a
 convenient metaphor
 employed to convey that at each
 instance of time the dynamical tide is modeled with a static tide where
 all the time-dependent variables are shifted back by $\,\Delta t\,$, i.e.,
 (a) the moon is rotated back by $\,\vbold\,\Delta t\,$, and (b) the
 attitude of the planet is rotated back by $\,\omegabold_p\,\Delta t\,$.
 From the viewpoint of a planet-based observer, this means that a dynamical
 response to a satellite located at $\,\erbold\,$ is modeled with a static
 response to a satellite located at $\,\erbold_f\,\equiv\,\erbold\,-\,
 \Delta t\,(\vbold\,-\,\omegabold_p\times\erbold)\,$.

 In this paper, we intend to dwell on geophysical issues -- the frequency-dependence of
 the attenuation rate and its consequences. Hence, to avoid unnecessary mathematical
 complications, in the subsequent illustrative example we shall restrict ourselves to the
 simple case of a tide-raising satellite on a near-equatorial near-circular orbit. In
 this approximation , the velocity of the satellite relative to the surface is
 \ba
 |\,\omegabold_p\times\erbold~-~\vbold\,|~=\;r\;|\,\omega_p\,-\,n\,|\;\;\;,
 \label{4}
 \label{404}
 \ea
 the principal tidal frequency is
 \ba
 \chi\;=\;2\;|\,\omega_p\,-\,n\,| \;\;\;\;,
 \label{5}
 \label{405}
 \ea
 and the angular lag is
 \ba
 \delta\;=\;\frac{\Delta t}{r}\;|\,\omegabold_p\times\erbold\;-\;
 \vbold\,|\;=\;\frac{\Delta t}{2}\;\chi\;\;\;,
 \label{6}
 \label{406}
 \ea
 $n\,$ being the satellite's mean motion, and $\,\omegabold_p\,$ being the planet's spin
 rate. The factor of two emerges in (\ref{5}) since the moon causes two elevations on the
 opposite sides of the planet. It will also be assumed that $\;\chi \Delta t\,\ll \,1\;$,
 for which reason we shall neglect the second-order difference between the expression
 (\ref{6}) and the angle subtended at the planet's centre between the moon and the tidal
 bulge (rigorously speaking, the sine of the subtended angle is equal to $\;\efbold\,
 \times\,\erbold/r^2\;$).

 The starting point of all tidal models is that each elementary volume of the planet is
 subject to a tide-raising potential, which in general is not periodic but can be expanded
 into a sum of periodic terms. Within the linear approximation introduced by Love,
 the tidal perturbations of the potential yield linear response of the shape and linear
 variations of the stress. In extension of the linearity approximation, it is always
 implied that the overall dissipation inside the planet may be represented as a sum of
 attenuation rates corresponding to each periodic disturbance:
 \ba
 \langle\,\dot{E}\;\rangle\;=\;\sum_{i}\;\langle\,
 \dot{E}(\chi_{\textstyle{_i}})\;\rangle
 \label{407}
 \ea
 where, at each frequency $\,\chi_i\,$,
 \ba
 \langle\,\dot{E}(\chi_{\textstyle{_i}})~\rangle~=~-~2~\chi_{\textstyle{_i}}~
 \frac{\,\langle\,E(\chi_{\textstyle{_i}})~\rangle\,}{Q(\chi_{\textstyle{_i}})
 }~=\,\;-\;\chi_{\textstyle{_i}}\;\frac{\,E_{_{peak}}(\chi_{\textstyle{_i}})
 \,}{Q(\chi_{\textstyle{_i}})}\;\;\;,
 \label{408}
 \ea
 $\langle\,.\,.\,.\,\rangle~$ standing for averaging over flexure cycle,
 $\,E(\chi_{\textstyle{_i}})\,$ denoting the energy of deformation at
 the frequency $\,\chi_{\textstyle{_i}}\,$, and $Q(\chi_{\textstyle{_i}})\,$
 being the quality factor of the material at this frequency. Introduced
 empirically as a means to figleaf our lack of knowledge of the attenuation
 process in its full complexity, the notion of $\,Q\,$ has proven to be practical
 due to its smooth and universal dependence upon the frequency and temperature.
 An alternative to employment of the empirical $\,Q\,$ factors would be
 comprehensive modeling of dissipation using a solution of the equations of
 motion, given a rheological description of the mantle (Mitrovica \& Peltier 1992;
 Hanyk, Matyska \& Yuen 1998, 2000; Moore \& Schubert 2000). Though providing a
 valuable yield for a geophysicist, this comprehensive approach may be avoided in
 astronomy, where only the final outcome, the frequency dependence of $\,Q\,$, is
 important. Fortunately, this dependence is already available from observations.

 In this paper we shall restrict ourselves to the simple case of an equatorial
 or near-equatorial satellite describing a circular or near-circular orbit. Under
 these circumstances only the principal tidal frequency (\ref{405}) will matter.

 \section{The quality factor $Q$ and the geometric lag angle $\delta $.}

 During tidal flexure, the energy attenuation through friction is, as ever,
 accompanied by a phase shift between the action and the response. The tidal
 quality factor is interconnected with the phase lag $\,\epsilon$ and
 the angular lag $\delta\,$ via
 \ba
 Q^{-1}\;=\;\tan \epsilon\;=\;\tan 2\delta
 \label{}
 \label{420}
 \ea
 or, for small lag angles,
  \ba
 Q^{-1}\;\approx\;\epsilon\;=\;2\;\delta\;\;\;.~~
 \label{}
 \label{421}
 \ea
 The doubling of the lag is a nontrivial issue. Many authors erroneously state that
 $\,Q^{-1}\,$ is equal simply to the tangent of the lag, with the factor of two
 omitted. For example, Rainey \& Aharonson (2006) assume that $\,Q^{-1}\,$ is equal
 to the tangent of the geometric lag. As a result, they arrive at a value of $\,Q\,$
 that is about twice larger than those obtained by the other teams. In Bills et al.
 (2005), one letter, $\,\gamma\,$, is used to denote two different angles. Prior to
 equation (24) in that paper, $\,\gamma\,$ signifies the {\emph{geometric}} lag
 (in our notations, $\,\delta_1\,$). Further, in their equations (24) and (25), Bills
 et al. employ the notation $\,\gamma\,$ to denote the {\emph{phase}} lag (in our
 notations, $\,\epsilon\,$, which happens to be equal to $\,2\,\delta_1\,$). With
 this crucial caveat, Bills' equation $\,Q\,=\,1/\tan\gamma\,$ is correct. This
 inaccuracy in notations has not prevented Bills et al. (2005) from arriving to a
 reasonable value of the Martian quality factor, $\,85.58\pm0.37\,$. (A more recent
 study by Lainey et al. (2007) has given a comparable value of $\,79.91 \pm 0.69\,$.)

 In the Appendix, we offer a simple illustrative calculation, which explains whence this
 factor of two stems.

 Formulae (\ref{420} - \ref{421}) look reasonable: the higher the quality factor, the
 lower the damping rate and, accordingly, the smaller the lag. What looks very far from
 being OK are the frequency dependencies ensuing from the assertions of $\,\delta\,$ being
 either constant or linear in frequency: the approach taken by Gerstenkorn (1955),
 MacDonald (1964), and Kaula (1964) implies that $\,Q\,\sim\,\chi^{0}\,$, while the theory
 of Singer (1968) and Mignard (1979, 1980) yields $\,Q\,\sim\,\chi^{-1}\,$, neither option
 being in agreement with the geophysical data.

 \section{Dissipation in the mantle.}

 \subsection{Generalities}

 Back in the 60s and 70s of the past century, when the science of low-frequency
 seismological measurements was yet under development, it was widely thought that
 at long time scales the quality factor of the mantle is proportional to the inverse
 of the frequency.
 This fallacy proliferated into planetary astronomy where it was received most
 warmly, because the law $\,Q\,\sim\,1/\chi\,$ turned out to be the only model for
 which the linear decomposition of the tide gives a set of bulges displaced from
 the direction to the satellite by the same angle. Any other frequency dependence
 $\,Q(\chi)\,$ entails superposition of bulges corresponding to the separate
 frequencies, each bulge being displaced by its own angle. This is the reason why
 the scaling law $\,Q\,\sim\,1/\chi\,$, long disproved and abandoned in geophysics
 (at least, for the frequency band of our concern), still remains a pet model in
 celestial mechanics of the Solar system.

 Over the past twenty years, considerable progress has been achieved in the
 low-frequency seismological measurements, both in the lab and in the field.
 Due to an impressive collective effort undertaken by
 several teams, it is now a firmly established fact that {\it{for
 frequencies down to about $\sim 1$ yr$^{-1}$ the quality factor of the mantle
 is proportional to the frequency to the power of a {\textbf{positive}} fraction}}
 $\,\alpha\,$. This dependence holds for all rocks within a remarkably broad band of
 frequencies: from several MHz down to about $\;1\;$yr$^{-1}\,$.

 At timescales longer than $\,1$ yr, all the way to the Maxwell time (about $\,100
 $~yr), attenuation in the mantle is defined by viscosity, so that the
 quality factor is, for all minerals, well approximated with $\,{\eta}\chi/M\,$,
 where $\eta$ and $M$ are the shear viscosity and the shear elastic modulus of the
 mineral. Although the values of both the viscosity coefficients and elastic moduli
 greatly vary for different minerals and are sensitive to the temperature, the
 overall quality factor of the mantle at such long timescales still is linear in
 frequency.

 At present there is no consensus in the seismological community in regard to the time
 scales exceeding the Maxwell time. One viewpoint (incompatible with the Maxwell
 model) is that the linear law $\,Q\,\sim\,\chi\,$ extends all the way down to the
 zero-frequency limit (Karato 2007). An alternative point of view (prompted by the
 Maxwell model) is that at scales longer than the Maxwell time we return to the
 inverse-frequency law $\,Q\,\sim\,1/\chi\,$.\\
 ~\\
 All in all, we have:\\
 \ba
 \mbox{For}\;\;\;\;\;\;10^7\;\;\mbox{Hz}\;\;>\,\;\chi\;\;>\;\;1\;yr^{-1}\;:\;\;\;
 Q\,\sim\,\chi^{\alpha}\;,\,\;\mbox{with}\;\,\alpha\,=\,0.2\,-\,0.4\;\;\;(\,0.2\;
 \mbox{for partial melts})\;.\;\;\;\,
 \label{one}
 \label{422}
 \ea
 \ba
 \mbox{For}\;\;\;\;\;\;\;\;\;1\;yr^{-1}\,>\,\chi\,>\,10^{-2}\,\;yr^{-1}\;\;:\;\;\;
 \;Q\,\sim\,\chi\;\;\;.\;\;\;\;~~~~~~~~~~~~~~~~~~~~~~~~~~~~~~~~~~~~~~~~~~~~~~~~~~~
 ~~~~~~~
 \label{two}
 \label{423}
 \ea
  \ba
 \mbox{For}\;\;\;\;\;10^{-2}\;yr^{-1}\,>\,\chi\;:\;\;\;\mbox{arguably, ~it ~is
 ~still~~}Q\,\sim\,\chi\;.\;\;\;\;\;(\,\mbox{Or~maybe~~~}Q\;\sim\;1/\chi~\,?\,)
 ~~~~~~~~~~~~~\,
 \label{three}
 \label{424}
 \ea
 Fortunately, in practical calculations of tides {\emph{in planets}} one never has
 to transcend the Maxwell time scales, so the controversy remaining in
 (\ref{three}) bears no relevance to our subject. We leave for a future study the
 case of synchronous satellites, the unique case of the Pluto-Charon resonance, or
 the binary asteroids locked in the same resonance. Thus we shall avoid also the
 frequency band addressed in (\ref{two}), but shall be interested solely in the
 frequency range described in (\ref{one}). It is important to emphasise that the
 positive-power scaling law (\ref{one}) is well proven not only for samples in the
 lab but also for vast seismological basins and, therefore, is universal. Hence,
 this law may be extended to the tidal friction -- validity of this extension will
 be discussed below in subsection 3.4.1

 Below we provide an extremely condensed review of the published data whence the
 scaling law (\ref{one}) was derived by the geophysicists. The list of sources will
 be incomplete, but a full picture can be obtained through the further references
 contained in the works to be quoted below. For a detailed treatment, see Chapter
 11 of the book by Karato (2007) that contains a systematic introduction into the
 theory of and experiments on attenuation in the mantle.

 \subsection{Circumstantial evidence: attenuation in minerals.\\
  Laboratory measurements and some theory}

 Even before the subtleties of solid-state mechanics with or without melt are
 brought up, the positive sign of the power $\,\alpha\,$ in the dependence $\,Q\,
 \sim\,\chi^{\alpha}\,$ may be anticipated on qualitative physical grounds. For a
 damped oscillator obeying $\;\ddot{z}\,+\,2\,\beta\,\dot{z}\,+\,\chi^2\,z\,=\,0\;
 \,$, the quality factor is equal to $\,\chi/(2\beta)\,$, i.e., $\,Q\,\sim\,\chi\,$.

 Solid-state phenomena causing attenuation in the mantle may be divided into
 three groups: the point-defect mechanisms, the dislocation mechanisms, and
 the grain-boundary ones.

 Among the point-defect mechanisms, most important is the transient diffusional
 creep, i.e., plastic flow of vacancies, and therefore of atoms, from one grain
 boundary to another. The flow is called into being by the fact that vacancies
 (as well as the other point defects) have different energies at grain boundaries
 of different orientation relative to the applied shear stress. This anelasticity
 mechanism is wont to obey the power law $\,Q\,\sim\,\chi^{\alpha}\,$ with
 $\,\alpha\,\approx\,0.5\,$.

 Anelasticity caused by dislocation mechanisms is governed by the viscosity law
 $\,Q\,\sim\,\chi\,$ valid for sufficiently low frequencies (or sufficiently high
 temperatures), i.e., when the viscous motion of dislocations is not restrained by
 the elastic restoring stress. (At higher frequencies or/and lower
 temperatures, the restoring force ``pins" the defects. This leads to the
 law $\,Q\,\sim\,(\textstyle{1\,+\,\tau^2\chi^2})\tau^{-1}\chi^{-1}\,$,
 parameter $\,\tau\,$ being the relaxation time whose values considerably
 vary among different mechanisms belonging to this group. As the mantle is
 warm and viscous, we may ignore this caveat.)

 The grain-boundary mechanisms, too, are governed by the law $\,Q\sim\chi^{\alpha}$,
 though with a lower exponent: $\,\alpha\approx 0.2 -0.3$. This behaviour gradually
 changes to the viscous mode ($\alpha=1$) at higher temperatures and/or at lower
 frequencies, i.e., when the elastic restoring stress reduces.

 We see that in all cases the quality factor of minerals should grow with frequency.
 Accordingly, laboratory measurements confirm that, within the geophysically
 interesting band of $\,\chi\,$, the quality factor behaves as $\,Q\,\sim\,
 \chi^{\alpha}\,$ with $\,\alpha\,=\,0.2\,-\,0.4\;$. Such measurements have been
 described in Karato \& Spetzler (1990) and Karato (1998). Similar results were
 reported in the works by the team of I. Jackson -- see, for example, the paper
 (Tan et al. 1997) where numerous earlier publications by that group are also
 mentioned.

 In aggregates with partial melt the frequency
 dependence of $\,Q\,$ keeps the same form, with $\,\alpha\,$ leaning to
 $\,0.2\,$ -- see, for example, Fontaine et al. (2005)
 and references therein.

 \subsection{Direct evidence: attenuation in the mantle.\\
 Measurements on seismological basins}

 As we are interested in the attenuation of tides, we should be prepared to face
 the possible existence of mechanisms that may show themselves over very large
 geological structures but not in small samples explored in the lab. No matter
 whether such mechanisms exist or not, we would find it safer to require that the
 positive-power
 scaling law $\,Q\,\sim\,\chi^{\alpha}\,$, even though well proven in the lab,
 must be propped up by direct seismological evidence gathered over vast zones
 of the mantle. Fortunately, such data are available, and for the frequency range
 of our interest these data conform well with the lab results. The low-frequency
 measurements, performed by different teams over various basins of the Earth's
 upper mantle, agree on the pivotal fact: the seismological quality factor scales
 as the frequency to the power of a {\emph{positive}} fraction $\,\alpha\;$ -- see,
 for example, Mitchell (1995), Stachnik et al. (2004), Shito et al. (2004), and
 further references given in these sources.\footnote{~So far, Figure 11 in Flanagan
 \& Wiens (1998) is the only experimental account we know of, which only partially
 complies with the other teams' results. The figure contains two plots depicting
 the frequency dependencies of $\,1/Q_{shear}\,$ and $\,1/Q_{compress}\,$. While
 the behaviour of both parameters remains conventional down to $\,10^{-1}\,$Hz,
 the shear attenuation surprisingly goes down when the frequency decreases to
 $\,10^{-3}\,$Hz. Later, one of the Authors wrote to us that ``{\emph{Both P and
 S wave attenuation becomes greater at low frequencies. The trend towards lower
 attenuation at the lowest frequencies in Fig. 11 is not well substantiated.}}"
 (D. Wiens, private communication) Hence, the consensus on (\ref{one}) stays.}

 \subsection{Consequences for the tides}

 \subsubsection{Tidal dissipation vs seismic dissipation}

 For terrestrial
 planets, the frequency-dependence of the $\,Q\,$ factor of bodily tides is similar
 to the frequency-dependence (\ref{one} - \ref{two}) of the seismological $\,Q\,$
 factor. This premise is based on the fact that the tidal attenuation in the mantle
 is taking place, much like the seismic attenuation, mainly due to the mantle's
 rigidity. This is a nontrivial fact because, in distinction from earthquakes,
 the damping of tides is taking place both due to nonrigidity and self-gravity of the
 planet. Modeling the planet with a homogeneous sphere of density $\,\rho\,$,
 rigidity $\,\mu\,$, surface gravity $\,\mbox{g}\,$, and radius $\,R\,$, Goldreich
 (1963) managed to separate the nonrigidity-caused and self-gravity-caused inputs into
 the overall tidal attenuation. His expression for the tidal quality factor has the
 form
 \ba
 Q\;=\;Q_o\left(\;1\;+\;\frac{2}{19}\;\frac{\mbox{g}\;\rho\;R}{\mu}\;\right)\;\;\;,
 \label{Goldreich}
 \ea
 $Q_o\,$ being the value that the quality factor would assume were self-gravity absent.
 To get an idea of how significant the
 self-gravity-produced input could be, let us plug there the mass and radius of Mars
 and the rigidity of the Martian mantle. For the Earth's mantle, $\,\mu\,=\,65\,\div\,
 80\,$GPa. Judging by the absence of volcanic activity over the past hundred(s)
 of millions of years of Mars' history, the temperature of the Martian upper mantle is
 (to say the least) not higher than that of the terrestrial one. Therefore we may safely
 approximate the Martian $\,\mu\,$ with the upper limit for the rigidity of the
 terrestrial mantle: $\,\mu\,=\,10^{11}\,$Pa. All in all, the relative
 contribution from self-gravity will look as
 \ba
 \frac{2}{19}\;\frac{\mbox{g}\;\rho\;R}{\mu}\;=\;
 \frac{6}{76\,\pi}\;\frac{\gamma\;M^2}{\mu\;R^4}\;
 \approx\;\frac{1}{40}\;\frac{(\;6.7\,\times\,10^{\textstyle{^{-11}}}\,
 \mbox{m}^{\textstyle{^3}}\,\mbox{kg}^{\textstyle{^{-1}}}\,\mbox{s}^{\textstyle{^{-2}}}\;)
 \;\;(\;6.4\,\times\,10^{\textstyle{^{23}}}\,\mbox{kg}\;)^{\textstyle{^2}}}{(\;
 10^{\textstyle{^{11}}}\;\mbox{Pa}\;)\;\;(\;3.4\,\times\,10^{\textstyle{^6}}\,m\;)^{
 \textstyle{^4}}}\;\approx\;5.2\,\times\,10^{\textstyle{^{-2}}}\;\;,\;\;\;\;
 \label{}
 \ea
 $\gamma$ denoting the gravity constant. This, very conservative estimate shows
 that self-gravitation contributes, at most, several percent into the
 overall count of energy losses due to tides.
 This is the reason why we extend to the tidal $\,Q\,$ the
 frequency-dependence law measured for the seismic quality factor.

 \subsubsection{Dissipation in the planet vs dissipation in the satellite}

 A special situation is tidal relaxation toward the state where one body shows
 the same side to another. Numerous satellites show, up to librations, the same
 face to their primaries. Among the planets, Pluto does this to Charon. Such a
 complete locking is typical also for binary asteroids. A gradual approach
 toward the synchronous orbit involves ever-decreasing frequencies, eventually
 exceeding the limits of equation (\ref{two}) and thus the bounds of the present
 discussion. Mathematically, this situation still may be tackled by means of
 (\ref{one}) until the tidal frequency $\,\chi\,$ decreases to $\;1\,$yr$^{-1}$,
 and then by means of (\ref{two}) while $\,\chi\,$ remains above the inverse
 Maxwell time of the
 planet's material. Whether the latter law can be extended to longer time scales
 remains an open issue of a generic nature that is not related to a specific
 model of tides or to a particular frequency dependence of $\,Q\;$. The generic
 problem is whether we at all may use the concept of the quality factor beyond
 the Maxwell time, or should instead employ, beginning from some low $\,\chi\,$,
 a comprehensive hydrodynamical model. In the current work, we address solely
 the satellite-generated tides on the planet. The input from the planet-caused
 tides on the satellite will be considered elsewhere. The case of Pluto will not
 be studied here either. Nor shall we address binary asteroids. (Since at present
 most asteroids are presumed loosely connected, and since we do not expect the
 dependencies (\ref{one} - \ref{two}) to hold for such aggregates, our theory
 should not, without some alterations, be applied to such binaries.)

 Thus, since we are talking only about dissipation inside the planet, and are
 not addressing the exceptional Pluto-Charon case, we may safely assume the
 tidal frequency to always exceed $\;1\,$yr$^{-1}$. Thence (\ref{one}) will
 render, for a typical satellite:
 \ba
 {\left.~~~~~~~~~\right.}
 Q\,\sim\,\chi^{\alpha}~~~,~~~~\mbox{with}~~\alpha~=~0.2~-~0.4~~~.
 \label{425}
 \ea
 Accordingly, (\ref{421}) will entail:
 \ba
 {\left.~~~~~~~~~\right.}
 \delta~\sim~\chi^{-\alpha}~~~,~~~~\mbox{with}~~\alpha~=~0.2~-~0.4~~~.
 \label{426}
 \ea


 Another special situation is a satellite {\emph{crossing}} a synchronous
 orbit. At the moment of crossing, the principal tidal frequency $\,\chi\,=
 \,2\,|\omega_p\,-\,n|\,$ vanishes. As (\ref{421}) and (\ref{one}) yield
 $\,Q\,\approx\,(2\delta)^{-1}\,$ and $\,Q\,\sim\,\chi^{\alpha}\,$, then we
 get $\,\delta\,\sim\,\chi^{-\alpha}\,$ with a positive $\,\alpha\,$.
 Uncritical employment of these formulae will then make one think that at
 this instant the lag $\,\delta\,$ grows infinitely, a clearly nonsensical
 result. The quandary is resolved through the observation that the bulge is
 lagging not only in its position but also in its height, for which reason
 the dissipation rate remains finite (Efroimsky 2007). Since in this
 paper we shall not consider crossing of or approach to synchronous orbits,
 and since the example we aim at is Phobos, we shall not go deeper into this
 matter here.

 \subsection{A thermodynamical aside: the frequency and the temperature}

 In the beginning of the preceding subsection we already mentioned that though the
 tidal $Q$ differs from the seismic one, both depend upon the frequency in the same
 way, because this dependence is determined by the same physical mechanisms. This
 pertains also to the temperature dependence, which for some fundamental reason
 combines into one function with the frequency dependence.

 As explained, from the basic physical principles, by Karato (2007, 1998), the
 frequency and temperature dependencies of $\,Q\,$ are inseparably connected. The
 quality factor can, despite its frequency dependence, be dimensionless only if it
 is a function not just of the frequency {\emph{per se}} but of a dimensionless product
 of the frequency by the typical time of defect displacement. This time exponentially
 depends on the activation energy $A^*$, whence the resulting function is
 \ba
 Q\;\sim\;\left[\,\chi\;\exp(A^*/RT)\,\right]^{\alpha}\;\;\;.
 \label{427}
 \ea
 For most minerals of the upper mantle, $A^*$ lies within the limits of
 $360-540$ kJ mol$^{-1}$. For example, for dry olivine it is about $520$
 kJ mol$^{-1}$.

 Thus, through formulae (\ref{427}) and (\ref{421}), the cooling rate of
 the planet plays a role in the orbital evolution of satellites: the lower
 the temperature, the higher the quality factor and, thereby, the smaller
 the lag $\, \delta\,$. For the sake of a crude estimate, assume
 that most of the tidal attenuation is taking place in some layer, for which
 an average temperature $T$ and an average activation energy $A^*$ may be
 introduced.
 Then from (\ref{427}) we have: $\,\Delta Q/{Q}\,\approx\,-\,\alpha\,A^*\,\Delta T/RT\,$.
 For a reasonable choice of values $\,\alpha=0.3\,$ and $\,A^*=\,5.4\,\times
 \,10^5\,J/mol\,$, a drop of the temperature from $\,T_o=2000\,K\,$ down by
 $\,\Delta T=200\,K\,$ will result in $\,\Delta Q/Q\,\approx\,1\,$. So a
 $\,10\,\%\,$ decrease of the temperature can result in an about $\,100\,\%\,$
 growth of the quality factor.

 Below we shall concentrate on the frequency dependence solely.

 \section{Formulae}

 The tidal potential perturbation acting on the tide-raising satellite
 is
 \ba
 U(\delta_1)\;=~\frac{A_2}{r_{\textstyle{_f}}^5\;r^5}\;\left(\,3\;\left(\erbold_f
 \cdot\erbold\right)^2\;-\;\erbold_f^{\;2}\;\erbold^{\;2}\,\right)\;+
 \;\frac{A_3}{r_{\textstyle{_f}}^6\;r^6}\;\left(\,5\;\left(\erbold_f
 \cdot\erbold\right)^2\;-\;3\;\erbold_f^{\;2}\;\erbold^{\;2}\,\right)
 \;+\;\;.\;\,.\;\,.~~
 \label{16}
 \ea
 where $\,r\,\equiv\,|\erbold|\,$ and $\,r_f\,\equiv\,|\erbold_f|\,$,
 while the constants are given by
 \ba
 A_2\;\equiv\;\frac{k_2\,G\,m\,R^5}{2}\;\;\;\;,\;\;\;\;\;\;A_3\;\equiv\;
 \frac{k_3\,G\,m\,R^7}{2}\;\;\;\;,\;\;\;\;\;\;\;.\,\;.\,\;.\;\;\;\;,
 \label{17}
 \ea
 $k_n\,$ being the Love numbers. (For derivation of (\ref{16}) see, for example, MacDonald
 (1964) and literature cited therein.)

 Three caveats will be appropriate at this point. First, to (\ref{16}) we should add the
 potential due to the tidal distortion of the moon by the planet. That input contributes
 mainly to the radial component of the tidal force exerted on the moon, and entails a
 decrease in eccentricity and semi-major axis (MacDonald 1964). Here we omit this term,
 since our goal is to clarify the frequency dependence of the lag. Second, we acknowledge
 that in many realistic situations the $\,k_3\,$ and sometimes even the $\,k_4\,$ term is
 relevant (Bills et al. 2005). With intention to keep these inputs in our subsequent work,
 here we shall restrict our consideration to the leading term
 only. Hence the ensuing formula for the tidal force will read:\footnote{~Be mindful that
 $\;O(\,{\efbold}^{\;\;{2}}/r^2)\;=\;O(\,\delta^2)\,\ll\,1\;$.}
 \ba
 \nonumber
 \efcal\;=\;-\;\frac{3\;k_2\;G\;m^2\;R^5}{r^{7}}\;\left[\;\frac{\erbold}{r}\;-
 \;\frac{\vec{\mbox{\it\textbf{f}}}}{r}\;-\;2\;\frac{\erbold}{r}\;
 \frac{\,\erbold\cdot{\vec{\mbox{\it\textbf{f}} }}\,}{r^2}\;+\;O(\,{\efbold}^{\;\;{2}}/r^2)
 \right]\;+\;O(k_3Gm^2R^7/r^{9})\;\;\;\\
 \nonumber\\
 \approx\;-\;\frac{3\;k_2\;G\;m^2\;R^5}{r^{10}}\;\left[\;{\erbold}\;{r^2}\;-
 \;{\vec{\mbox{\it\textbf{f}}}}\;{r^2}\;-\;2\;{\erbold}\;\left(\,\erbold
 \cdot{\vec{\mbox{\it\textbf{f}}}}\,\right)\;\right]\;\;\;,\;\;\;\;\;\;\;\;\;\;\;\;
 \;\;\;\;~~~~~~~~~~~~~~~~~~~~~~
 \label{18}
 \ea
 The third important caveat is that in our further exploitation of this formula we
 shall take into account the frequency-dependence of the lag $\,\fbold\,$, but not
 of the parameter $\,k_2\,$. While the dependence $\,\fbold(\chi)\,$ will be derived
 through the interconnection of $\,\fbold\,$ with $\,\delta(\chi)\,$ and therefore with
 $\,Q(\chi)\,$, the value of $\,k_2\,$ will be asserted constant. That the latter is
 acceptable can be proven through the following formula obtained by Darwin (1908)
 under the assumption of the planet being a Maxwell body (see also Correia \& Laskar
 2003):
 \ba
 \nonumber
 k_2(\chi)\;=\;k_{\textstyle{_{fluid}}}\;\,\sqrt{\;\frac{1\;+\;\chi^2\;\eta^2/\mu^2}{1\;+\;
 \left(\,\chi^2\;\eta^2/\mu^2\,\right)\;\left(\,1\;+\;19\;\mu/(2\,g\,\rho\,R)\,\right)^2}\;}
 \label{k_2}\;\;\;.
 \ea
 Here $\,k_{\textstyle{_{fluid}}}\,$ is the so-called fluid Love number. This is the
 value that $\,k_2\,$ would have assumed had the planet consisted of a perfect fluid
 with the same mass distribution as the actual planet. Notations $\,\mu\,$,
 $\,\rho\,$, $\,\mbox{g}\,$, and $\,\mbox{g}\,$ stand for the rigidity, mean density,
 surface gravity, and the radius of the planet. For these parameters, we shall keep
 using the estimates from subsection 3.4. The letter $\,\eta\,$ signifies the
 viscosity. Up to an order or two of magnitude, its value may be approximated, for a
 terrestrial planet's mantle, with $\,10^{22}\,$kg/(m$\cdot$ s). This will yield:
 $\,\chi^2\;\eta^2/\mu^2\,=\,(\,\chi\,\cdot\,10^{11}\,\mbox{s}\,)^2\,$,
 wherefrom we see that in all realistic situations pertaining to terrestrial planets
 the frequency-dependence in Darwin's formula will cancel out. Thus we shall neglect
 the frequency-dependence of the Love number $\,k_2\,$ (but shall at the same time
 take into account the frequency-dependence of $\,Q\,$, for it will induce
 frequency-dependence of all three lags).

 The interconnection between the position, time, and angular lags,
 \ba
 \delta\;\equiv\,\;\frac{|{\vec{\mbox{\it\textbf{f}}}} \,|}{r}\,\;=\;\Delta t\;\;
 \frac{1}{r}\;|\;\omegabold_p\times\erbold\;-\;\vbold\;|\;\,=\;\frac{\Delta t}{2}
 \;\chi~~~,
 \label{19}
 \ea
 can be equivalently rewritten as:
 \ba
 {\vec{\mbox{\it\textbf{f}}}}~=~{\bf{\hat{f}}}\;r\;\delta\;=\;r\;\frac{\Delta t}{2}
 \;\chi\;{\bf{\hat{f}}}\;\;\;,
 \label{20}
 \ea
 where
 \ba
 {\bf{\hat{f}}}\;=\;\frac{~\omegabold_p \times \erbold\,-\,\vbold}{~|\,
 \omegabold_p\times\erbold\,-\,\vbold\,|}
 \label{21}
 \ea
 is a unit vector pointing in the lag direction.

 Be mindful that we assume the inclination and eccentricity to be small, wherefore
 the ratio
 \ba
 \nonumber
 \frac{|{\vec{\mbox{\it\textbf{f}}}} \,|}{r}\,\;=\;\Delta t\;\;
 \frac{1}{r}\;|\;\omegabold_p\times\erbold\;-\;\vbold\;|
 \ea
 is simply the tangential angular lag, i.e., the geometric angle subtended at
 the primary's centre between the moon and the bulge. In the general case of a finite
 inclination or/and eccentricity, all our formulae will remain in force, but the lag $\,
 \delta\,$ will no longer have the meaning of the subtended angle.

 At this point, it would be convenient to introduce a dimensional integral parameter
 $\,{\cal{E}}\,$ describing the overall tidal attenuation rate in the planet. The
 power scaling law mentioned in section 3 may be expressed as
 \ba
 Q\,=\,{\cal E}^{\alpha}\,\chi^{\alpha}~~~,~~~
 \label{Q}
 \ea
 where $\,{\cal E}^{\alpha}\,$ is simply the dimensional factor
 emerging in the relation $\,Q\,\sim\,\chi^{\alpha}\,$. As mentioned in subsection
 3.5, cooling of the planet should become a part of long-term orbital calculations.
 It enters these calculations through evolution of this parameter $\,{\cal{E}}\,$.
 Under the assumption that most of the tidal dissipation is taking place in some
 layer, for which an average temperature $T$ and an average activation energy $A^*$
 may be defined,
 (\ref{427}) yields:
 \ba
 \nonumber
 {\cal E}\;=\;{\cal E}_o\;\exp\left[\,\frac{A^*}{R}\;
 \left(\,\frac{1}{T_o}\;-\;\frac{1}{T}\;\right)\;\right]\;\;\;,
 \label{}
 \ea
 $T_o\,$ being the temperature of the layer at some fiducial epoch. The physical
 meaning of the integral parameter $\,{\cal{E}}\,$ is transparent: if the planet
 were assembled of a homogeneous medium, with a uniform temperature distribution,
 and if attenuation in this medium were caused by one particular physical mechanism,
 then $\,{\cal{E}}\,$ would be a relaxation time scale associated with this
 mechanism (say, the time of defect displacement). For a realistic planet,
 $\,{\cal{E}}\,$ may be interpreted as a relaxation time averaged (in the sense of
 $\,Q\,=\,{\cal E}^{\alpha}\,\chi^{\alpha}\,$) over the planet's layers and over
 the various damping mechanisms acting within these layers.

 As (\ref{one}) entails $\;\delta\,\approx\,1/(2\,Q)\,=\,(1/2)\,
 {\cal{E}}^{-\alpha}\,\chi^{-\alpha}\;$, then (\ref{19}) necessitates for the position lag:
 \ba
 {\vec{\mbox{\it\textbf{f}}}}~=~r\;\delta\;{\bf{\hat{f}}}\;=\;\frac{1}{2}\;
 r\;{\cal E}^{-\alpha}\;\chi^{-\alpha}\;{\bf{\hat{f}}}\;\;\;,
 \label{23}
 \ea
 and for the time lag:
 \ba
 {\Delta t}\;=\;{\cal E}^{-\alpha}\;\chi^{-(\alpha +1)}\;\;\;,
 \label{22}
 \ea
 ${\cal E}\,$ being the planet's integral parameter introduced above, and $\,\chi\,$ being
 a known function (\ref{5}) of the orbital variables. Putting everything together, we
 arrive at
 \ba
 {\vec{\mbox{\it\textbf{f}}}}~=\,\frac{1}{2}\,\left({\cal{E}}{\chi}\right)^{-\alpha}\,a\;
 \,\frac{1\,-\,e^2}{1\,+\,e\;\cos\nu}\;\,\frac{~\omegabold_p \times\erbold\,-\,\vbold~}{~|
 \,\omegabold_p\times\erbold\,-\,\vbold\,|~}~=~\frac{1}{2}\,\left({\cal{E}}{\chi}
 \right)^{-\alpha}\,a\;\,\frac{~\omegabold_p \times\erbold\,-\,\vbold~}{~|
 \,\omegabold_p\times\erbold\,-\,\vbold\,|~}\;+\;O(e)\;\;,~~
 \label{24}
 \ea
 where
 \ba
 \chi\,\equiv\;2\;|\omega_p\,-\;n|\;\;.\;\;
 \label{25}
 \ea
 The time lag is, according to (\ref{22}):
 \ba
 \Delta
 t=
 {\cal{E}}\;\left(\,2\;{\cal{E}}\;|\omega_p\,-\;n|\,
 \right)^{\textstyle{^{-\,(\alpha+1)}}}.~
 \label{39}
 \ea
 Formulae (\ref{39}), (\ref{24}), and (\ref{18}) are sufficient to both compute the orbit
 evolution and trace the variations of the time lag.

 \section{The example of Phobos' fall to Mars}

 As an illustrative example, let us consider how the realistic dependence
 $\,Q(\chi)\,$ alters the life time left for Phobos. We shall neglect the
 fact that Phobos is close to its Roche limit, and may be destroyed by
 tides prior to its fall. We also shall restrict the dynamical interactions
 between Phobos and Mars to a two-body problem disturbed solely with the
 tides raised by Phobos on Mars. Thus we shall omit all the other
 perturbations, like the Martian non-sphericity and precession, or the pull
 exerted upon Phobos by the Sun, the planets, and Deimos. If, along with
 these simplifications, we assume the eccentricity and inclination to be
 small, then we shall be able to describe the evolution of the semi-major axis
 by means of the following equation (Kaula 1964, p. 677, formula 41): \be
 \frac{da}{dt}\;=\;-\;\frac{3\;k_2\;R^5\;G\,m}{Q\;\sqrt{G\,(M_o\,+\,m)}\;
 a^{11/2}}\;\;\;,
 \label{eq:Kaula}
 \ee
 with $\,M_o\,$ and $\,m\,$ denoting the masses of Mars and Phobos. This equation
 can be solved analytically, provided the quality factor $\,Q\,$ is set constant
 (as in Kaula 1964). The solution is:
 \be
 a(t)\;=\;\left(\;-\;\frac{39\;k_2\;R^5\;G\,m}{2\;Q\sqrt{G\,(M\,+\,m)}}\;t
 \;+\;a_o^{13/2}\right)^{2/13}
 \label{solution}
 \ee
 $a_o\,\equiv\,a(t)|_{\textstyle{_{t=0}}}\,$ being the initial value of Phobos'
 semi-major axis.

 Unfortunately, neither our model (wherein $\,Q\,$ is given by (\ref{Q})$\,$) nor
 the Singer-Mignard model (with $\,Q\,$ scaling as $\,1/\chi\;$) admit such
 an easy analytical solution. This compels us to rely on numerics. In a
 (quasi)inertial frame centered at Mars, the equation of motion looks:
 \be
 \frac{d^2\erbold}{dt^2}\;=\;-\;\frac{G\,(M_o\,+\,m)\;\erbold}{r^3}\;-\;
 \frac{3\;k_2\;G\,(M_o\,+\,m)\;m\;R^5}{r^{10}\;M_o}
 \;\left[\;-\;{\vec{\mbox{\it\textbf{f}}}}\;r^2\;-\;2\;\erbold\;
 \left(\;\erbold\cdot{\vec{\mbox{\it\textbf{f}} }}\,\right)\;
 \right]
 \label{eq:phobosXYZ}
 \ee
 Naturally, its right-hand side consists of the principal, two-body
 contribution $\;-\,{G(M_o+m)\erbold}/{r^3}\,$ and the disturbing tidal
 force given by (\ref{18}). It should be noted, however, that we did
 not bring in here all the terms from (\ref{18}). Following Mignard (1980), we
 retain in (\ref{eq:phobosXYZ}) only the perturbing terms dependent on
 $\,{\vec{\mbox{\it\textbf{f}}}}\,$. The other perturbing term (the first term
 on the right-hand side of (\ref{18})$\,$) is missing in (\ref{eq:phobosXYZ}),
 because it provides no secular input into the semi-major axis' evolution.
 For a proof of this fact see Appendix A.3 below.

 Phobos' orbital motion obeys the planetary equations in the Euler-Gauss form.
 In assumption of $\,\inc\,$ and $\,e\,$
 %
 %
 %
 %
 being small, the problem conveniently reduces to one equation:
 \be
 \frac{da}{dt}=\frac{2S}{n}\;\;\;,
 \label{eq:gauss}
 \ee
 where $\,S\,$ is the tidal acceleration given by the second term of
 (\ref{eq:phobosXYZ}) projected onto the tangential direction of the satellite
 motion. This direction is defined by the unit vector
 $\;{(\Hbold\times\erbold)}/{|\Hbold\times\erbold|}\;$, where $\,\Hbold\,$
 denotes the angular-momentum vector. Thence the said projection
 reads:
 \ba
 S\;=\;-\;\frac{3\;k_2\;G\;(M_o\,+\,m)\;m\;R^5}{r^{10}\;M_o}\;\left[\;-\;r^2
 \;\Delta t\;\left(\;\omegabold_p\,\times\,\erbold\;-\;\vbold\;\right)\;+\;
 2\;\erbold\;\Delta t\;(\erbold\,\cdot\,\vbold)\,\right]\,\cdot\,\frac{(\Hbold
 \times\erbold)}{|\Hbold\times\erbold|}\;\;\;.\;\;\;\;\;
 \label{projection}
 \ea
 In Mignard (1980) the appropriate expression is given with a wrong sign. This
 is likely to be a misprint, because the subsequent formulae in his
 paper are correct.

 In assumption of $\,\inc\,$ and $\,e\,$ being negligibly small,
 $\,(\Hbold\,\times\,\erbold)\cdot(\,\omegabold_p\,\times\,\erbold\,)\,$ can
 be approximated with $\,n\,a^4\,\omega_p\,$, whereafter (\ref{projection}) gets
 simplified to
 \be
 S\;=\;-\;\frac{3\;k_2\;R^5\;G\;(M_o\,+\,m)\;m\;\Delta t}{M_o\;a^7}\;\left(n\;-\;
 \omega_p\right)\;\;\;,
 \label{eq:S}
 \ee
 substitution whereof into (\ref{eq:gauss}) entails the following
 equation to integrate:
 \be
 \frac{da}{dt}\;=\;-\;\frac{6\;k_2\;R^5\;n\;m\;\Delta t}{M_o\;a^4}\;\left(n\;-\;
 \omega_p\right)\;\;\;.
 \label{eq:Gauss-Simplify}
 \ee
 %
 Our computational scheme was based on the numerical integrator RA15 offered by
 Everhart (1985).
 The initial value of $\,a\,$, as well as the values of all the other
 physical parameters entering (\ref{eq:Gauss-Simplify}), were borrowed from
 Lainey et al. (2007). These included an estimate of $\,0.6644 $ $min$ for
 the present-day time lag $\,\Delta t\,$.

 Four numerical simulations were carried out. One of these implemented the
 Singer-Mignard model with a tidal-frequency-independent $\,\Delta t\,$.
 The other three integrations were performed for the realistic frequency-dependence
 (\ref{39}), with $\,\alpha\,=\,0.2\,$, $0.3$ and $0.4$.

 To find the integral parameter $\,{\cal E}\,$ emerging in
 (\ref{22}), we used the present-day values of $\,\Delta t\,$ and $\,n\,$.
 The resulting values of $\,{\cal E}\,$ were found to be $\,1201\,\times\,10^5$
 $day\;rad^{-1}\,$, $\,81028\,~day\;rad^{-1}$ and $\,2104\,$ $day\;rad^{-1}\;$
 for $\,\alpha\,=\,0.2\,$, $0.3\,$ and $\,0.4\,$, respectively. We did not take
 into account the strong temperature-dependence of $\,{\cal E}\,$, leaving this
 interesting topic for discussion elsewhere.

\begin{figure}
 \includegraphics[width=12.1cm,angle=-90]{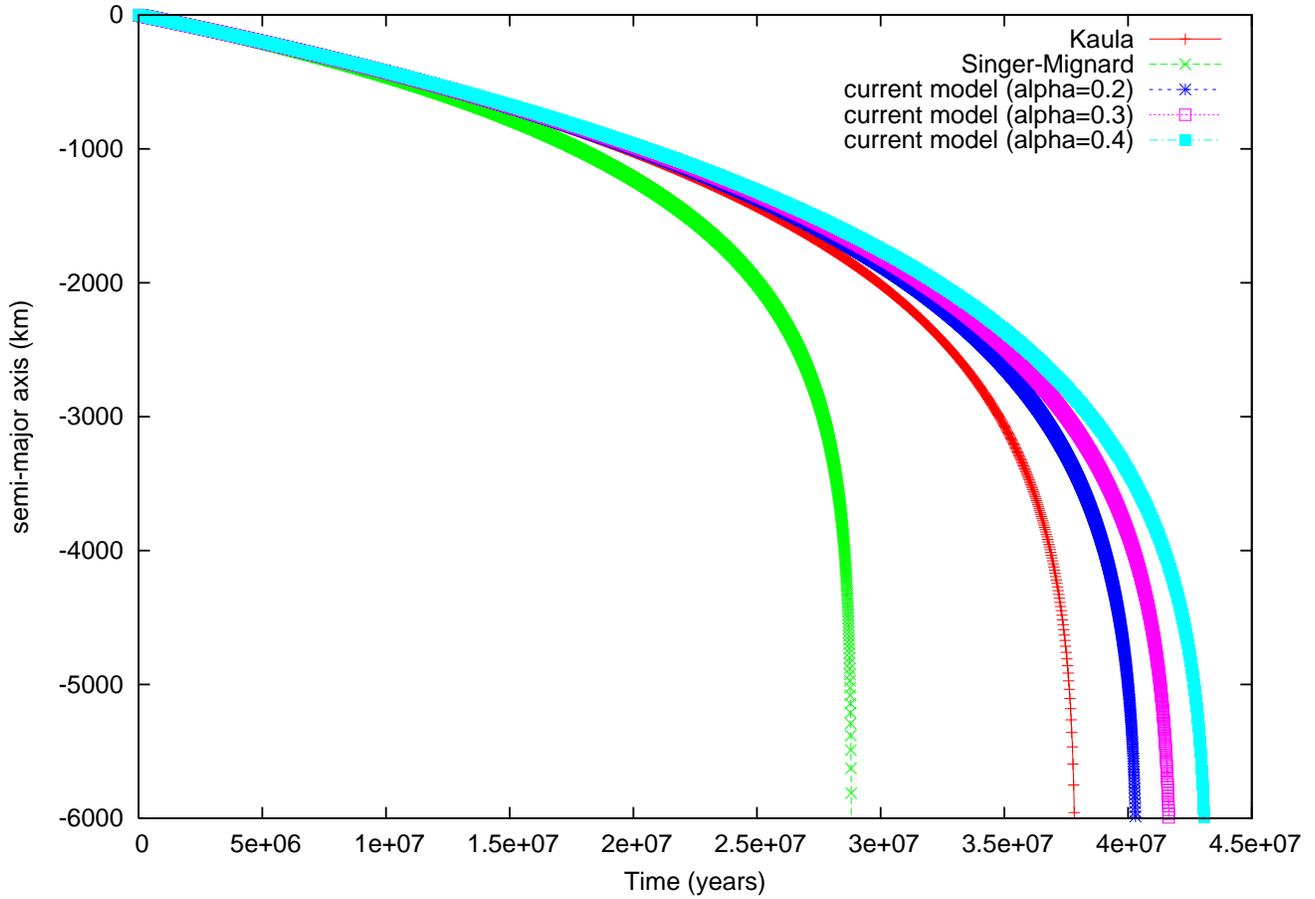}

 \caption{\small ~~Evolution of Phobos' semi-major axis, as predicted by
 different models. The lines (from left to right) correspond to the
 Singer-Mignard model, to the Gerstenkorn-McDonald-Kaula model, and to the
 realistic rheology with $\,\alpha\,=\,0.2\,$, with $\,\alpha\,=\,0.3\,$,
 and with $\,\alpha\,=\,0.4\,$, correspondingly.}
  \label{Efroimsky-Lainey-fig1.eps}
  \end{figure}

\begin{figure}
 \includegraphics[width=12.1cm,angle=-90]{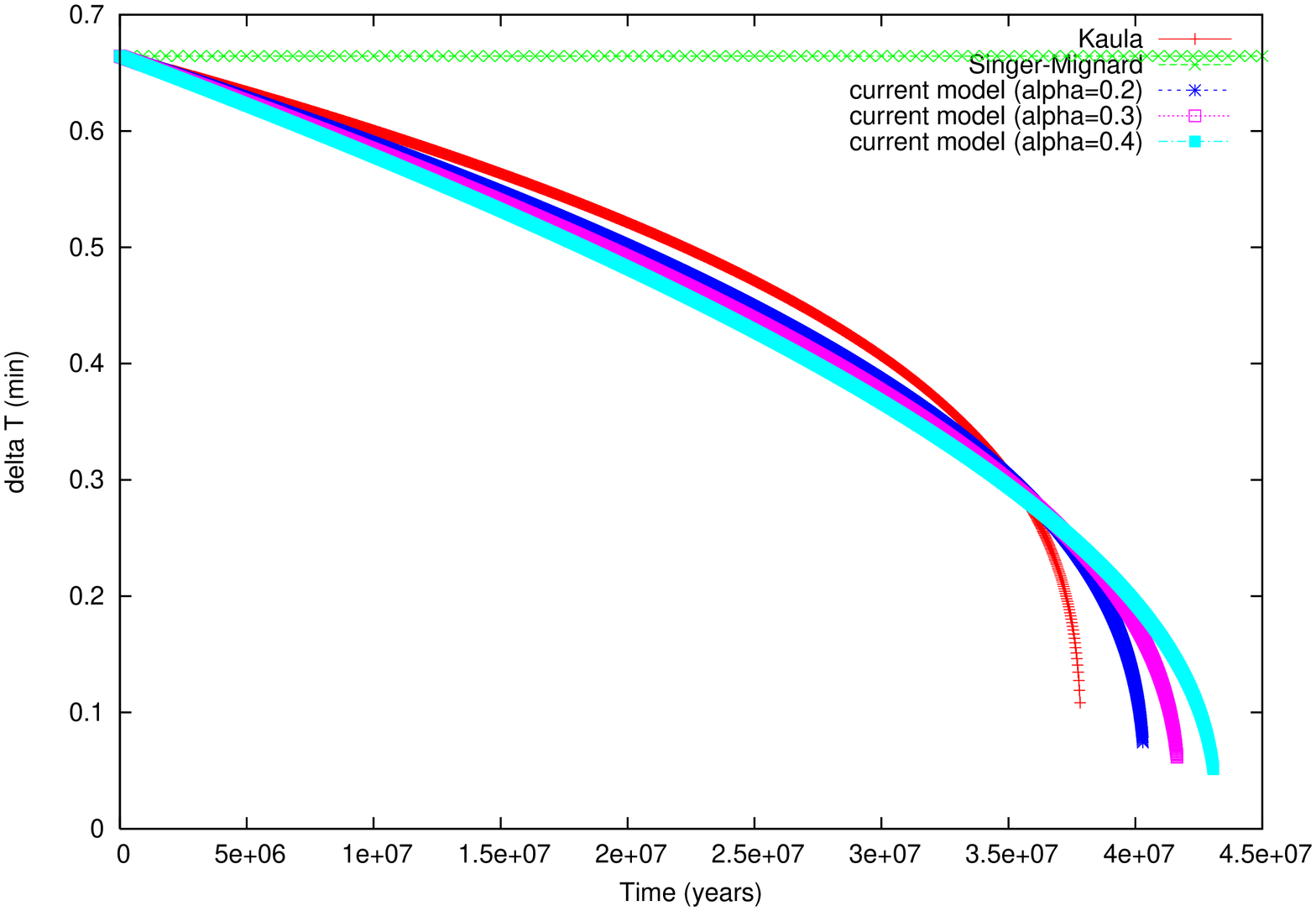}

 \caption{\small ~~Evolution of $\Delta t$ computed for three different
 models. The horizontal green line corresponds to the Singer-Mignard model
 wherein $\,\Delta t\,$ is set to be constant. The red line corresponds
 to the Gerstenkorn-MacDonald-Kaula model. The dark-blue, violet, and
 light-blue lines correspond to the realistic rheological models with
 $\,\alpha\,=\,0.2\;$, with $\,\alpha\,=\,0.3\,$. and with $\,\alpha\,=\,0.4\,$,
 correspondingly.}

  \label{Efroimsky-Lainey-fig2.eps}
  \end{figure}

 Simultaneous numerical integration of equations (\ref{39}) and
 (\ref{eq:Gauss-Simplify}) results in plots presented on Figures
 \ref{Efroimsky-Lainey-fig1.eps} and \ref{Efroimsky-Lainey-fig2.eps}.
 The first of these pictures shows the evolution of Phobos' semi-major axis
 from its present value until the satellite crashes on Mars, having
 descended about $\,6000\,$ km. The leftmost curve reproduces the known
 result that, according to the Singer-Mignard model with a constant $\,\Delta t\,$,
 Phobos should fall on Mars in about $\,29\,$ Myr.\footnote{~In the paragraph after
 his formula (18), Mignard (1981) states that ``Phobos will end its life in about
 $\,36\,$ million years". Mignard arrived to that number by using an old estimate of
 $\,20\,$ deg/cyr$^2$ for the initial tidal acceleration. Later studies, like for
 example Jacobson et al. (1989) and Lainey et al. (2007) have shown that this value
 should be increased to $\,25.4\,$ deg/cyr$^2$. It is for this reason that our
 simulation based on the Singer-Mignard model gives not $\,36\,$ but only $\,29\,$
 Myr for Phobos' remaining lifetime.}
 The next curve was obtained not numerically but analytically. It depicts the
 analytical solution (\ref{solution}) available for the Gerstenkorn-MacDonald-Kaula
 model with a constant $\,Q\,$, and demonstrates that this model promises to Phobos a
 longer age, $\,38\,$ Myr. The three curves on the right were obtained by numerical
 integration of (\ref{39}) and (\ref{eq:Gauss-Simplify}). They correspond to the
 realistic rheology with $\,\alpha\,$ equal to $\,0.2\,$, $\,0.3\,$ and $\,0.4\,$. It
 can be seen that within the realistic model Phobos is expected to survive for about
 $\,40\,-\,43\,$ Myr, dependent upon the actual value of $\,\alpha\,$ of the Martian
 mantle. This is about $\,15\,$ Myr longer than within the Singer-Mignard model
 widely accepted hitherto.

 The difference between the three scenarios shown on Figure
 \ref{Efroimsky-Lainey-fig1.eps} stems from the different rate of evolution
 of the lag $\,\Delta t\,$ in the three theories addressed. Within the
 Singer-Mignard formalism, $\,\Delta t\,$ stays unchanged through the
 descent. As can be seen from formula (\ref{39}), this is equivalent to
 setting $\,\alpha\,=\,-\,1\,$, an assertion not supported by geophysical
 data. Within the Gerstenkorn-MacDonald-Kaula model, the time lag is subject
 to a gradual decrease described by the formula
 \be
 \Delta t\;=\;\frac{\arctan(1/Q)}{2\;|n\,-\,\omega_p|}
 \label{dt}
 \ee
 under the assumption that $\,Q\,$ is constant and is equal to its present-day
 value $\,Q\,=\,79.91\,$ determined in (Lainey et al. 2007). Comparison of
 (\ref{dt}) with (\ref{39}) reminds us of the simple fact that, in terms of our
 model, Gerstenkorn-MacDonald-Kaula's theory corresponds to the choice of
 $\,\alpha\,=\,0\,$, a choice which is closer to the realistic rheology than
 the Singer-Mignard model.

 In the realistic model, $\,\alpha\,$ is positive and assumes a value of about
 $\,0.2\,-\,0.4\,$. As a result, the time lag is gradually decreasing. However
 this decrease looks different from that in Gerstenkorn-MacDonald-Kaula's model
 -- for their comparison see Figure~\ref{Efroimsky-Lainey-fig2.eps}.

 \section{Conclusions}

 As the tidal angular lag $\,\delta\,$ is inversely proportional to the
 tidal $\,Q\,$ factor, the actual frequency-dependence of both $\,\delta\,$
 and $\,\Delta t\,$ is unambiguously defined by the frequency-dependence of
 $\,Q\,$. While in the Gerstenkorn-MacDonald-Kaula theory of tides the geometric
 lag is assumed frequency-independent, in the Singer-Mignard theory it is the time
 lag that is spared of frequency dependence. However, neither of these two choices
 conform to the geophysical data.

 We introduce a realistic tidal model, which permits the quality factor and,
 therefore, both the angular lag $\,\delta\,$ and the time lag $\,\Delta
 t\,$ to depend on the tidal frequency $\,\chi\,$. The quality factor is
 wont, according to numerous studies, to obey the law $\,Q\,\sim\,
 \chi^{\alpha}\,$, where $\,\alpha\,$ lies within $\,0.2\,-\,0.4\,$.
 This makes the time lag $\,\Delta t\,$ not a constant but a function
 (\ref{22}) of the principal tidal frequency and, through (\ref{39}), of the
 orbital elements of the satellite. The same pertains to the angular lag
 $\delta$.

 Using these tidal-frequency dependencies for the time and angular lags,
 along with the recently updated values of the Martian parameters, we
 explored the future of Phobos, taking into account only the tides raised
 by Phobos on Mars, but not those caused by Mars on Phobos. Our integration
 shows that Phobos will fall on Mars in $\,40\,-\,43\,$ Myr from now. It
 is up to $\,50\,\%\,$ longer than the estimate stemming from the
 Singer-Mignard model employed in the past. This demonstrates that the currently
 accepted time scales of dynamical evolution, deduced from old tidal models,
 should be reexamined using the actual frequency dependence of the lags.


 ~\\

 {\underline{\textbf{\Large{Acknowledgments}}}}\\
 ~\\
 ME would like to deeply thank Peter Goldreich, Francis Nimmo, William Moore,
 and S. Fred Singer for their helpful comments and recommendations. VL wishes
 to gratefully acknowledge his fruitful conversation with Attilio Rivoldini
 concerning dissipation in Mars. The authors' very special gratitude goes to
 Shun-ichiro Karato whose consultations on the theory and phenomenology of the
 quality factor were crucially important for the project.

 ~\\


 {\underline{\textbf{\Large{Appendix.}}}}\\
 ~\\
 The goal of this Appendix is threefold. First, we remind the reader why in
 the first approximation the quality factor is inversely proportional to the
 phase lag. Second, we explain why the phase lag is twice the geometric lag
 angle, as in formulae (\ref{420} - \ref{421}) above. While a comprehensive
 mathematical derivation of this fact can be found elsewhere (see the unnumbered
 formula between equations (29) and (30) on p. 673 in Kaula 1964), here
 we illustrate this counterintuitive result by using the simplest setting.
 Third, we justify our neglect of the first term in (\ref{18}).
 ~\\
 ~\\
 \noindent
 {\textbf{{A.1.~~The case of a near-circular near-equatorial orbit.}}}\\
  ~\\
 Consider the simple case of an equatorial moon on a circular orbit. At
 each point of the planet, the tidal potential produced by this moon will read
  \ba
  W\;=\;W_o\;\cos \chi t\;\;\;,
  \label{A3}
  \label{469}
  \ea
 the tidal frequency being given by
 \ba
 \chi\,=\,2~|n\;-\;\omega_p|~~~.~~~
 \label{A3}
 \label{470}
 \ea
 Let $\,\mbox{g}\,$ denote the free-fall acceleration. An element of the planet's volume
 lying beneath the satellite's trajectory will then experience a vertical elevation of
 \ba
 \zeta\;=\;\frac{W_o}{\mbox{g}}\;\cos (\chi t\;-\;2\delta)\;\;\;.
 \label{A4}
 \label{471}
 \ea
 Accordingly, the vertical velocity of this element of the planet's volume will
 amount to
  \ba
 u\;=\;\dot{\zeta}\;=\;-\;\chi\;\frac{W_o}{\mbox{g}}\;\sin (\chi t
 \;-\;2\;\delta)\;=\;-\;\chi\;\frac{W_o}{\mbox{g}}\;\left(\sin \chi
 t\;\cos 2\delta\;-\;\cos \chi t\; \sin 2\delta\right)\;\;.\;\;
 \label{A5}
 \label{472}
 \ea
 The expression for the velocity has such a simple form because in this case the
 instantaneous frequency $\,\chi\,$ is constant. The satellite generates two
 bulges -- on the facing and opposite sides of the planet -- so each point of the
 surface is uplifted twice through a cycle. This entails the factor of two in the
 expressions (\ref{470}) for the frequency. The phase in (\ref{A4}), too, is
 doubled, though the necessity of this is less evident.\footnote{~Let $\,x\,$
 signify a position along the equatorial circumference of the planet. In the
 absence of lag, the radial elevation at a point $\,x\,$ would be:
 \ba
 \nonumber
 \zeta\;=\;\frac{W_o}{\mbox{g}}\;\cos
 k(x\;-\;v\;t)\;\;\;,\;\;\;\;\;\;v\,=\,R\,\sigma\;\;\;,
 \ea
 $v\,$ being the velocity of the satellite's projection on the ground,
 $\,R\,$ being the planet's radius, and
 $\sigma$ being simply $\,|n-\omega_p|\,$ because we are dealing with a
 circular equatorial orbit. The value of $\,k\,$ must satisfy
 \ba
 \nonumber
 k\;v\;=\;2\;\sigma\;\;\;,\;\mbox{i.e.,}\;\;\;\;k\;v\;=\;\chi\;\;\;,
 \ea
 to make sure that at each $\,x\,$ the ground elevates twice per
 an orbital cycle. The above two formulae yield:
 \ba
 \nonumber
 k\;R\;=\;2\;\;\;.
 \ea
 In the presence of lag, all above stays in force, except that the
 formula for radial elevation will read:
 \ba
 \nonumber
 \zeta\;=\;\frac{W_o}{\mbox{g}}\;\cos k(x\;-\;v\;t\;+\;D)\;\;\;,\;\;\;\;
 \mbox{where}\;\;\;\; D\;=\;R\;\delta\;\;\;,
 \ea
 $D\,$ being the linear lag, and $\,\delta\,$ being the angular
 one. Since $\,k\,v\,=\,2\,$, we get:
 \ba
 \nonumber
 \cos \left[\,k\;(x\;-\;v\;t\;+\;R\;\delta_{_1})\,\right]\;=\;\cos \left[\,k\;x
 \;-\;k\;v\;t\;+\;k\;R\;\delta\,\right]\;=\;\cos
 \left[\,k\;x\;-\;(k\;v\;t\;-\;2\;\delta)\,\right]\;\;\;,
 \ea
 so that, at some fixed point (say, at $\,x\,=\,0\,$) the elevation becomes:
 \ba
 \nonumber
 \zeta(t)\;=\;\frac{W_o}{\mbox{g}}\;\cos (k\;v\;t\;-\;2\;\delta)\;\;\;.
 \ea
 We see that, while the geometric lag is $\;\delta\,$, the phase lag is
 double thereof.}

 The energy dissipated over a time cycle $\,T\,=\,2\pi/\chi\,$, per
 unit mass, will, in neglect of horizontal displacements, be
 \ba
 \nonumber
 \Delta E_{_{cycle}} &=& \int^{T}_{0}u\left(-\,\frac{\partial W}{
 \partial r}\right)dt=
 \,-\left(-\,\chi \frac{W_o}{\mbox{g}}\right)\,\frac{\partial W_o}{
 \partial r}\int^{t=T}_{t=0}\cos \chi t\,\left(\sin \chi t\,
 \cos 2\delta\,-\,\cos \chi t\, \sin 2\delta\right)dt\\
 \nonumber\\
 \nonumber\\
 &=&\,-\;\chi\;\frac{W_o}{\mbox{g}}\;\frac{\partial W_o}{\partial r}\;\sin 2\delta
 \;\frac{1}{\chi}\;\int^{\chi t\,=\,2\pi}_{\chi t\,=\,0}\;\cos^2 \chi t\;\;d(\chi
 t)\;=\;-\;\frac{W_o}{\mbox{g}}\;\frac{\partial W_o}{\partial r}\;\pi\;\sin2\delta
 \;\;,\;\;\;~~~~~~~~~~~~~~~~~~~~
 \label{A6}
 \label{}
 \ea
 while the peak energy stored in the system during the cycle will read:
 \ba
 \nonumber
 E_{_{peak}}&=&\int^{T/4}_{0} u \left(-\,\frac{\partial W}{
 \partial r}\right)dt =
 \,-\left(-\,\chi\,\frac{W_o}{\mbox{g}}\right)\frac{\partial W_o
 }{\partial r}\int^{t=T/4}_{t=0}\cos \chi t\,\left(\sin
 \chi t\,\cos 2\delta\,-\,\cos \chi t\,\sin 2\delta\right)dt\\
 \nonumber\\
 \nonumber\\
 \nonumber
 &=&\;2\;\sigma\;\frac{W_o}{\mbox{g}}\;\frac{\partial W_o}{\partial r}\;\left[\;
 \frac{\cos 2\delta}{\chi}\;\int^{\chi t\,=\,\pi/2}_{\chi t\,=\,0}
 \;\cos \chi t\;\sin \chi t\;\;d(\chi t)\;-\;\frac{\sin 2\delta
 }{\chi}\;\int^{\chi t\,=\,\pi/2}_{\chi t\,=\,0}\;\cos^2 \chi t
 \;\;d(\chi t)\;\right]\\
 \nonumber\\
 \nonumber\\
 &=&\;\frac{W_o}{\mbox{g}}\;\frac{\partial W_o}{\partial r}\;\left[\;\frac{1}{2}
 \;\cos 2\delta\;-\;\frac{\pi}{4}\;\sin 2\delta\;\right]~~~~~~~~~~~~~~~~~~~
 ~~~~~~~~~~~~~~~~~~~~~~~~~~~~~~~~~~~~~~~~~~~~~~~~~~~~~~~
 \label{A7}
 \label{}
 \ea
 whence

 \pagebreak

 \ba
 Q^{-1}\;=\;\frac{-\;\Delta E_{_{cycle}}}{2\,\pi\,E_{_{peak}}}\;=\;\frac{1}{2\,\pi}
 \;\,\frac{\pi\;\sin 2\delta}{~\frac{\textstyle 1}{\textstyle 2}\;\cos2\delta\;-\;
 \frac{\textstyle\pi}{\textstyle 4}\;\sin 2 \delta}\;\approx\;\tan 2 \delta\;\;\;.
 \label{A8}
 \label{}
 \ea
 The above formulae were written down in neglect of horizontal displacements,
 approximation justified below in the language of continuum mechanics.
 ~\\
 ~\\
 {\textbf{{A.2.~~On the validity of our neglect of the horizontal displacements}}}\\

 In our above derivation of the interrelation between $\,Q\,$ and $\,\delta\,$, we
 greatly simplified the situation, taking into account only the vertical
 displacement of the planetary surface, in response to the satellite's pull. Here
 we shall demonstrate that this approximation is legitimate, at least in the case
 when the planet is modeled with an incompressible and homogeneous medium.

 As a starting point, recall that the tidal attenuation rate within a tidally
 distorted planet is well approximated with the work performed on it by the
 satellite\footnote{~A small share of this work is being spent
 for decelerating the planet rotation}:
 \ba
 \dot{E}\;=\;-\;\int\,\rho\;\Vbold\;\nabla W\;d^3x
 \label{A9}
 \ea
 where $\,\rho\,,\;\Vbold\,$ and $\,W\,$ are the density, velocity, and tidal
 potential inside the planet. To simplify this expression, we shall employ the equality
 \ba
 \rho\,\Vbold\,\nabla W\,=\,\nabla\,\cdot\,(\rho\;\Vbold\;W)\;-\;W\;\Vbold\cdot
 \nabla\rho\;-\;W\;\nabla\,\cdot\,(\rho\;\Vbold)\,=\,\nabla\,\cdot\,(\rho\;
 \Vbold\;\nabla W)\;-\;W\;\Vbold\cdot\nabla\rho\;+\;W\;
 \frac{\partial\rho}{\partial t}\;\;.\;\;\;
 \label{}
 \ea
 For a homogeneous and incompressible primary, both the $\,\Vbold\;W \nabla
 \rho\,$ and $\,\partial\rho/\partial t\,$ terms are nil, wherefrom
 \ba
 \dot{E}\;=\;-\;\int\rho\;W\;\Vbold\,\cdot\,{\vec{\bf{n}}}\;d^3x\;\;\;,
 \label{}
 \ea
 ${\vec{\bf{n}}}\,$ being the outward normal to the surface of the planet. We
 immediately see that, within the hydrodynamical model, it is only the radial
 elevation rate that matters.

 Now write the potential as $\;W\,=\,W_o\,\cos(\chi\,t)\;$. Since the response
 is delayed by $\, \Delta t\,$, the surface-inequality rate will evolve as $\;
 \Vbold\cdot{\vec{\bf{n}}}\,\sim\,\sin\left[\,\chi\,(\,t\,-\,\Delta t\,)\,\right]
 \;$. All the rest will then be as in subsection A.1 above.

 ~\\
 {\textbf{{A.3.~~On the validity of our neglect of the nondissipative tidal
 potential}}}\\

 The right-hand side of equation (\ref{eq:phobosXYZ})
 consists of the principal part, $\;-\,{G\,(M_o\,+\,m)\,\erbold}/{r^3}\,$, and
 tidal perturbation terms. These are the second and third terms from the right-hand
 side of (\ref{18}), terms that bear a dependence on $\,\efbold\,$ and, therefore,
 on $\,\Delta t\,$. The first term from the right-hand side of (\ref{18}) lacks such
 a dependence and, therefore, is omitted in (\ref{eq:phobosXYZ}).
 The term was dropped because it would provide no secular input into the history of
 the semi-major axis. Here we shall provide a proof of this statement.

 The omitted term corresponds to a potential (Mignard 1980):
 \begin{eqnarray}
 U_0\;=\;\frac{k_2\;(M_o\,+\,m)\;G\;m\;R^5}{2\;M_o\;r^6}\;\;\;.
 \end{eqnarray}
 From the physical standpoint, $\,U_0\,$ models the effect of the
 tidal bulges, assuming their direction to coincide with that toward
 the tide-raising satellite. This potential entails no angular-momentum
 exchange, and therefore yields no secular effect on the semi-major axis.
 To prove this, let us decompose this potential into a series over the powers
 of $\,e\,$. This will require of us to derive the expression of $\,\left(
 {\textstyle{a}}/{\textstyle{r}}\right)^6\,$. Starting out with the well
 known development
 \begin{eqnarray}
 \nonumber
 \frac{a}{r}&=&1+\sum_{p=1}^\infty 2J_p(pe)\cos(pM)\\
 \nonumber\\
 \nonumber
 &=&1+e\cos(M)+e^2\cos(2M)+\left(\frac{9}{8}\cos(3M)-\frac{1}{8}\cos(M)\right)e^3\\
 \nonumber\\
 &+&\left(-\frac{1}{3}\cos(2M)+\frac{4}{3}\cos(4M)\right)e^4+\left(-\frac{81}{128}
 \cos(3M)+\frac{1}{192}\cos(M)+\frac{625}{384}\cos(5M)\right)e^5 \nonumber\\
 \nonumber\\
 &+&\left(\frac{81}{40}\cos(6M)+\frac{1}{24}\cos(2M)-\frac{16}{15}\cos(4M)\right)
 e^6+...~~~,~~~~~~~~~~~~~~~~~~~~~~~~~~~~~~~~~~~~~~~~~~~~~~~~~~~~
 \end{eqnarray}
 one can arrive to the following expansion:
 \begin{eqnarray}
 \nonumber
 \left(\frac{a}{r}\right)^6&=&1+6e\cos(M)+\left(\frac{15}{2}+\frac{27}{2}\cos(2M)
 \right)e^2+\left(\frac{107}{4}\cos(3M)+\frac{117}{4}\cos(M)\right)e^3\\
 \nonumber\\
 \nonumber
 &+&\left(\frac{197}{4}\cos(4M)+\frac{101}{2}\cos(2M)+\frac{105}{4}\right)e^4 \\
 \nonumber\\
 \nonumber
 &+&\left(\frac{5157}{64}\cos(3M)+\frac{5529}{64}\cos(5M)+\frac{2721}{32}\cos(M)
 \right)e^5\\
 \nonumber\\
 &+&\left(\frac{4839}{40}\cos(4M)+129\cos(2M)+\frac{525}{8}+\frac{732}{5}\cos(6M)
 \right)e^6+... ~~~,
 \end{eqnarray}
 whose average over the mean anomaly looks like:
 \begin{eqnarray}
 \langle\;\,\left(\frac{a}{r}\right)^6\;\rangle&=&1+\frac{15}{2}e^2+\frac{105}{4}
 e^4+\frac{525}{8}e^6+...
 \end{eqnarray}
 Hence, the averaged potential will become:
 \begin{eqnarray}
 \langle\,U_0\,\rangle=\frac{k_2(M_0+m)GmR^5}{2M_0a^6}\left(1+\frac{15}{2}e^2+
 \frac{105}{4}e^4+\frac{525}{8}e^6+...\right)
 \end{eqnarray}
 In the case of Phobos, the terms of order $\,O(e^2)\,$ may, in the first
 approximation, be neglected. This means that out of the six Lagrange-type
 planetary equations the first five will, in the first order of $\,e\,$, stay
 unperturbed, and therefore the elements $\,a\,,\,e\,,\omega\,,\,\inc\,,\,\Omega
 \,$ will, in the first order over $\,e\,$, remain unchanged. The Lagrange
 equation for the longitude will be the only one influenced by $\,U_0\,$.
 That equation will assume the form:
 \begin{eqnarray}
 \frac{dL}{dt}\;=\;n\;-\;\frac{2}{n\;a}\;\,\frac{\partial\;\langle\,U_0\,\rangle}{
 \partial \,a}
 \end{eqnarray}
 which gives
 \begin{eqnarray}
 L\;=\;n\;t\;+\;\frac{6\;k_2\;(M_o\,+\,m)\;G\;m\;R^5}{M_o\;n\;a^8}\;t
 \end{eqnarray}
 We see that in the first order of $\,e\,$ the only secular effect stemming from
 the potential $\,U_0\,$ is a linear in time evolution of the longitude.



 {}

\end{document}